# Accelerated Ions from a Laser Driven Z-pinch

Michael H. Helle[1, a)], Daniel F. Gordon, Dmitri Kaganovich, Yu-hsin Chen[2)], John P. Palastro[1)], and Antonio Ting

[1]U.S. Naval Research Laboratory, 4555 Overlook Ave., SW, Washington, DC 20375
[2]Research Support Instruments, 4325-B Forbes Boulevard, Lanham, MD 20706
[a)]Corresponding author: mike.helle@nrl.navy.mil

**Intense laser acceleration of ions is inherently difficult due to the velocity mismatch between laser pulses moving at the speed of light and slowly moving massive ions. Instead of directly accelerating the ions, current approaches rely on TV/m laser fields to ionize and drive out electrons. The ions are then accelerated by the resulting electrostatic fields from charge separation. Here we report experimental and numerical acceleration of ions by means of laser driven Z-pinch exiting a sharp plasma interface. This is achieved by first driving a plasma wakefield in the self-modulated bubble regime. Cold return currents are generated to maintain quasi-neutrality of the plasma. The opposite current repel and form an axial fast current and a cylindrical-shell cold return current with a large (100 MG) azithmuthal field in between. These conditions produce a Z-pinch that compresses the fast electrons and ions on axis. If this process is terminated at a sharp plasma interface, a beam of ions are then accelerated in the forward direction as the Z-pinch collapses. These results provide a new route towards laser-accelerated ions as well as potential laboratory scale modeling of z-pinches in astrophysical events.**

High-energy ions have been accelerated by means of laser interactions with over-dense or tenuous plasmas for over a decade[1,2]. These investigations relied on varying accelerating mechanisms, the most prominent being Laser Hole Boring[3,4], Target Normal Sheath Acceleration (TNSA), and Radiation Pressure Acceleration (RPA). All of these mechanisms primarily rely on accelerating fields produced by charge separation between the massive ions and laser accelerated hot electrons. While progress continues to be made, these experiments have relied on solid targets that can limit repetition rate and purity of the accelerated ions.

An alternative approach is to use the fields produced by a large current electron beam exiting a sharp plasma density boundary. As the electron beam propagates through the plasma, a cold electron return current is formed to balance the fast current and maintain plasma quasi-neutrality[5,6]. The two oppositely signed currents repel one another, forming an axial fast current and a cylindrical-shell cold return current. Simultaneously, a large (100 MG) azimuthal magnetic field is generated within the region between the two opposite currents. The mobile ions and electrons then pinch on-axis to generate the pressure gradients necessary to balance the JxB force on the plasma. This process is commonly known as a Z-pinch. For a short pulse fast electron beam at the exit of the plasma region, the situation can be setup where the rapid removal of the JxB force causes the pinched ions to explode. This process combined with the axial space charge fields between the electron beam and the ions further accelerates the ions in the beam direction. When driven by an intense laser, this results in a process with characteristics similar to the magnetic vortex acceleration mechanism observed in simulations [7-12]. To date, unambiguous signatures of such an acceleration process have yet to be observed and most of the work has been limited to the effects of the magnetic fields generated in 2D simulations.

In order to investigate the ions accelerated by this process, we used the electron beam generated by a laser wakefield accelerator (LWFA). Our experimental set-up is shown in figure 1(a). The goal of most LWFA experiments is to generate quasi-monoenergetic high-energy (100's MeV to 10's GeV) for various applications. For these experiments, we wish to drive a high current density electron beam and thus a mildly relativistic, broad energy spread electron beam is suitable. To achieve this we operated in the self-modulated bubble regime at near critical plasma densities. In previous experiments, we observed that the self-modulated bubble regime enabled reliable injection and acceleration of a high charge electron beam [13]. Operating at near critical densities is beneficial in that it: encourages injection through a low plasma wakefield phase velocity[14], produces a large current density due to small bubble radius ($r_B=2\sqrt{(a_0)}/k_p$)[15], and reduced the ion response time, $1/\omega_{pi}=(4\pi n_i Z^2 e^2/m_i)^{-1/2}$ where $k_p$ is the k-vector associated with the plasma wave, $a_0$ is the normalized vector potential, $n_i$ is the ion density, Z is the charge state, e is the unit of charge, and $m_i$ is the mass of the ion.

In order to reach the high densities required to initiate this process as well as produce the sharp density gradients to couple out the ions, we used a density tailoring technique recently developed at NRL[12,16]. Using two blastwaves generated by nanosecond frequency doubled Nd:YAG pulses, the hydrogen gas flow from a supersonic gas jet is hydrodynamically shaped into a 75 μm thick gas "foil" with 30 μm gradients and tunable peak densities from $2.5*10^{20}$ cm$^{-3}$-$5*10^{20}$ molecules cm$^{-3}$. This provides a pure hydrogen target at densities and gradients previously unexplored. The formation of the target is monitored using a 400nm, 50fs laser probe that goes to shadowgraphic and interferometric diagnostics. This is shown in figure 1(b).

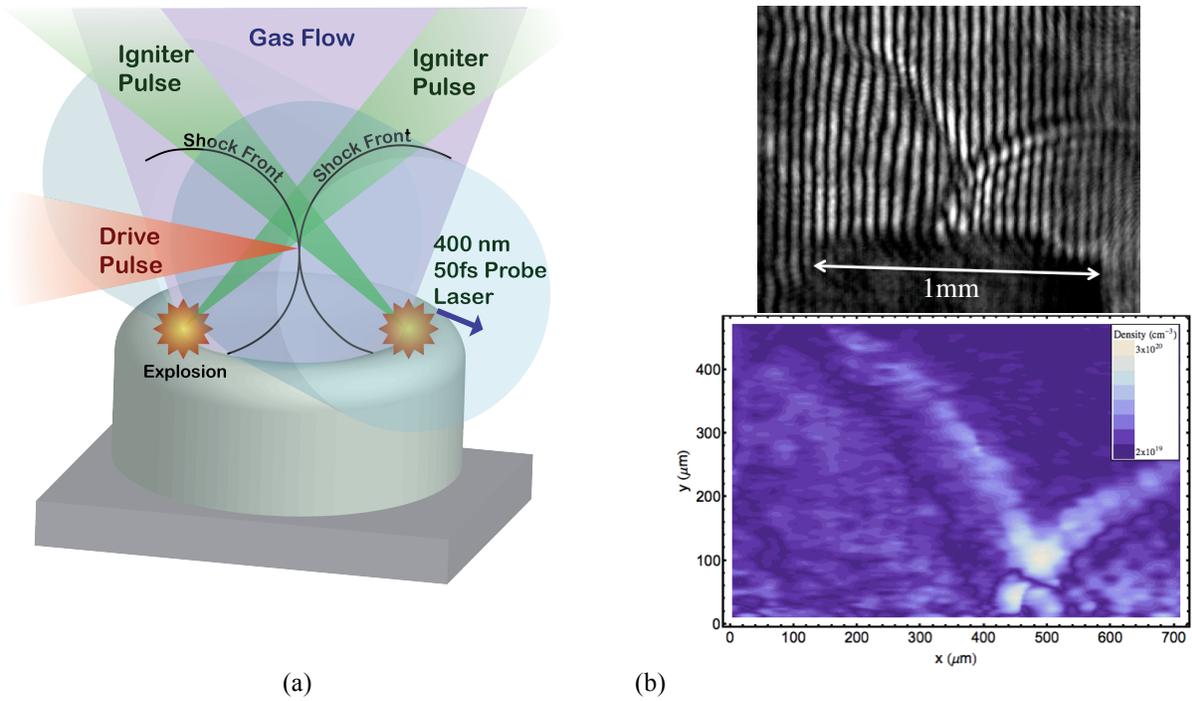

(a)                                      (b)

**FIGURE 1.** A schematic of the experimental setup including the 50fs probe (a) is on the left while an interferometric image of the target area (b) and the extracted density profile (c) are given on the right. A 1mm supersonic gas jet delivers the neutral gas while laser ignited strong hydrodynamic shocks act to shape it. The case when two colliding shocks are used to produce a ~70μm target is shown.

The interaction is driven by a 500 mJ, 800 nm, 50 fs laser pulse generated by the TFL laser system at NRL. The pulse is focused, using an *f*/2 off-axis parabola (OAP), to a vacuum spot size of 2.6 μm, $1/e^2$, reaching a peak intensity of $1*10^{20}$ W cm$^{-2}$. At this intensity the laser pulse is above the threshold for reaching the bubble regime. We focused the beam at the front edge of the "foil" to allow the beam to relativistically self-focus within the density ramp. This allows the beam to reach higher intensities and better match the bubble radius at the peak density. The energy and spatial distribution of the ions is measured using a 1 mm thick CR-39 plate with aluminum filters of varying thickness spaced perpendicular to the laser polarization axis. The stack was placed 13 cm from the gas jet to prevent laser damage of the plate.

A density scan was performed over the operational range of our gas jet at intervals of $0.5*10^{20}$ molecules cm$^{-3}$. Assuming fully ionized hydrogen, this corresponds to peak plasma densities of $n_e = 0.3\text{-}0.6\ n_{crit}$, where $n_{crit} = (2\pi c)^2 m_e\varepsilon_0/e^2\lambda_0^2$, $m_e$ is the mass of the electron and $\lambda_0$ is the laser wavelength. It was observed that at densities above $0.4\ n_{crit}$, a proton beam contained within the length of detector was accelerated in the forward direction. The beam was unable to penetrate even a single filter layer (9 μm) and thus had peak energy of <700 keV. Upon closer examination of the tracks produced, the proton energy is <200 keV [Methods]. An example detector image is given in Fig. 2 (a). At a density of $0.3\ n_{crit}$, we observed protons penetrating through the 9μm, 18μm, and 27μm thick filters with no proton penetrating the 36μm thick filter (figure 2(b)). This indicates a broad energy distribution of protons with a maximum energy between 1.5-1.9 MeV. Unlike the higher density cases, the spatial distribution of the protons within the unshielded region extends beyond the length of the detector. It is characterized by an intense beam near the left bound with a highly nonuniform distribution to the right of that. These uniformities are consistent with the distribution of higher energy protons observed in the shielded regions above and below, and appear to be one half of the high-energy beam generated by the exploding pinch.

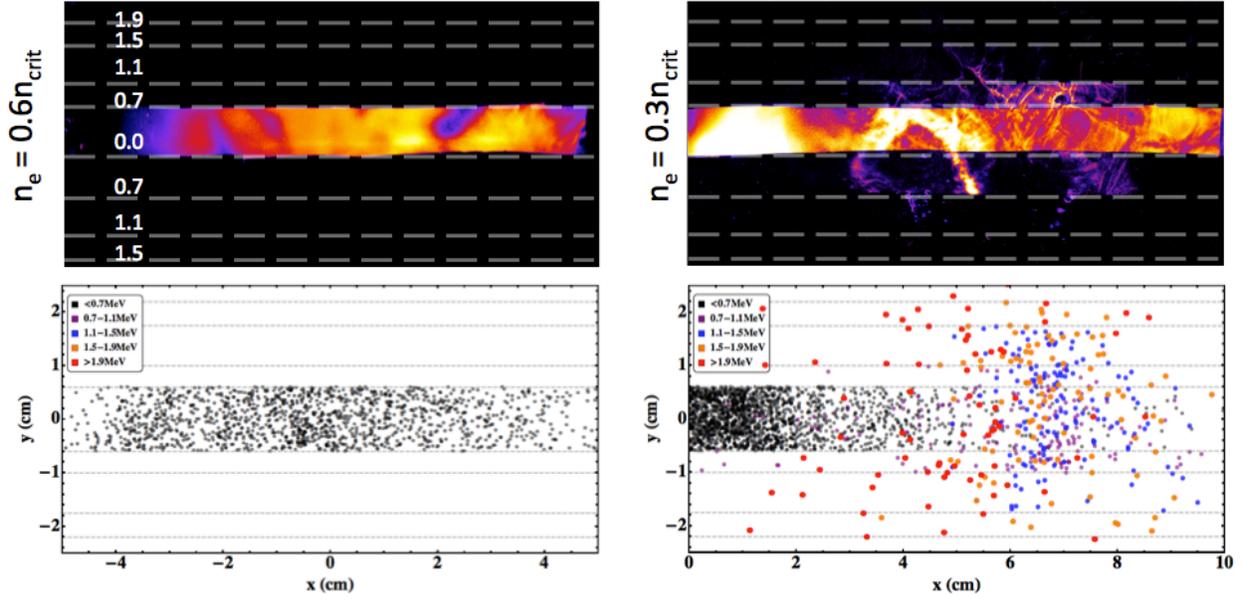

**FIGURE 2.** False color image of a 10cm x 5cm CR-39 plates for the cases of $0.6\ n_{crit}$ and $0.3\ n_{crit}$. The dashed lines represent the location of the filters and their lower energy bound is listed on the left. On the bottom are plots of simulated test particles projected out to the plane of the detector color-coded to the experimental energy buckets.

The origin of these accelerated protons was investigated using the TurboWAVE 3D particle-in-cell code [Methods]. The simulation was initialized using the experimental parameters and density profiles. Test particles were placed within the simulation box, their orbits were tracked, and projected to the detector plane. The projected ion test particles for the cases of $0.6n_{crit}$ and $0.3n_{crit}$ are given in Figs. 2 and 3. In figure 2(b) the test particle distributions are filtered to match the experimental conditions and show excellent agreement. To understand the underlying process, the full simulated distribution with the corresponding proton density profiles are given in figure 3. For the $0.6n_{crit}$ case, the protons are contained within a beam width of 8 cm FWHM and have a peak energy of 140 keV, consistent with the experimental track diameters [Methods]. Examining the results of the PIC simulation, we see that the laser pulse is unable to penetrate the target, figure 3b. The protons that are accelerated in the forward direction originate from the rear side of the target. They are accelerated by space-charge fields produced from hot electrons generated by the laser-target interaction exiting the rear of the target. This is consistent with the Target Normal Sheath Acceleration mechanism (TNSA) [17].

Examining the $0.3n_{crit}$ case, we see a vastly different situation. The forward accelerated protons are characterized by a low energy beam on-axis with a halo of high-energy protons with energies <2MeV. When examining the proton density plot, we observe that at these densities the laser pulse is able to penetrate through the entire plasma. The Z-pinch is also clearly seen extending through the plasma. To better illustrate the dynamic process that is occurring, 3D renderings showing the evolution of the electron plasma density, magnetic field lines, and proton test particles are given in figure 4. Initially, the intense laser pulse undergoes self-focusing and modulation instability in the density up ramp of the plasma region. The pondermotive force of the pulse drives electron cavitations. Plasma electrons become trapped and accelerated within these cavitations. Due to the slow plasma wave phase velocity and short dephasing length, electrons are easily injected, accelerated, and then stream out of the trapping region of the wakefield. These fast electrons trail behind the laser front, forming an axial fast current. To maintain plasma quasi-neutrality, a cold electron return current is formed to balance the fast current. The two oppositely signed currents repel one another, forming an axial fast current and a cylindrical-shell cold return current. The on-axis fast current has an average current of ~22kA. This process leads to the generation of a large (~50MG) azimuthal magnetic field contained within the region. During this process, the mobile ions pinch on-axis and an electron-ion pinch is formed with $n_e \approx n_i \approx .3 n_{crit}$ and radius of ~1μm. This whole process trails the intense laser pulse until it exits the plasma region.

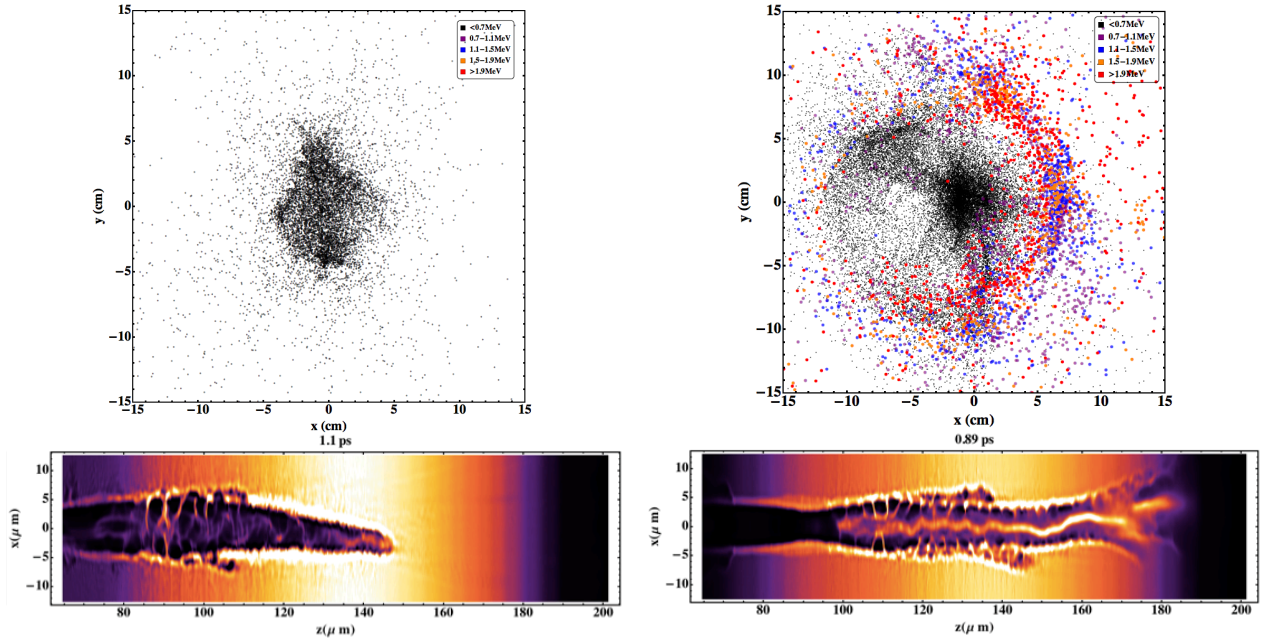

**FIGURE 3.** Plots of the full test particles spatial distribution as projected out to the plane of the detector for $0.6 n_{crit}$ and $0.3 n_{crit}$. Below are proton density plots taken along the polarization plane for each case. Note that for the $0.6 n_{crit}$ case the cavitation region extends to 150μm while it extends the entire length of the plasma for $0.3 n_{crit}$.

At the exit, the laser pulse ponderomotively expels the ambient electrons from the density ramp. After the intense laser pulse exits the plasma, the fast electron current and large azimuthal magnetic fields begin to flow out, and a situation similar to the magnetic vortex acceleration mechanism occurs[7-12]. The protons at the interface undergo the same inward pinch as those inside the main plasma region, however the magnitude of pinch rapidly drops off due to the drop in density. The fast current eventually leaves the plasma since it has a longitudinal extent comparable to the plasma region. At this point the magnetic fields begin to dissipate and the protons, no longer confined, explode outward radially. At the same time they acquire longitudinal momentum from the space-charge fields setup by the escaping fast current. These protons are those observed in the high-energy halo. Additionally, there exist a low-density population of protons outside the ramp. These are accelerated collectively by the fast electrons leaving the plasma region. These protons are the source of low energy protons on axis. Examining the test particle energy in the final rendering, it can be seen that the proton beam exhibits an energy distribution that is radially dependent as in the experiments.

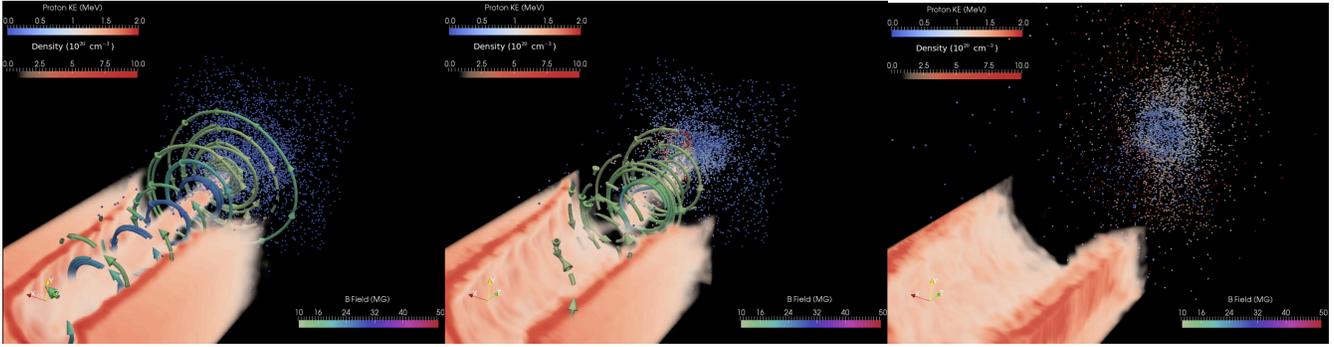

**FIGURE 5.** 3D volumetric renderings of the electron density and azimuthal fields with proton test particles included. The test particles are color coded by their kinetic energy. Note that test particles have a radial energy distribution with lowest energy on-axis and the highest off-axis.

Based on this model the acceleration provided by the electron beam should be entirely longitudinal, while the transverse temperature of the ions originating within the density ramp should come primarily from the pinch process. The divergence of the high-energy halo should therefore be directly related to the temperature of the ions in the pinch. Experimentally, the divergence angle of the highest energy protons (1.5-1.9MeV) in the .33$n_{crit}$ case is 6.5$^o$ This gives an ion temperature of $k_B T_i$ = 250-300keV. We can then compare this to value to that obtained using well-known pinch scaling. To lowest-order, the ion beam temperature is dictated by the Bennent condition for a Z-pinch, namely

$$I^2 \mu_0/4\pi = 2 \tilde{n} k_B (T_e + Z T_i)$$

Here I is the electron beam current, ñ is the number of electrons per unit length in the laboratory frame, $k_B$ is the Boltzman's constant, and T is the average electron and ion temperature. To calculated the electron beam current and electrons per unit length, we use the wakefield scalings presented in [14,15]. The electron beam temperature is calculated using previous experimental results[Methods]. A detailed discussion of the calculation is presented in the methods sections, however solving for $k_B T_i$ we obtain a transverse temperature of ~300keV for the protons. This compares favorably to the experimental values obtained above as well as to the average thermal energy of 310keV extracted from the corresponding simulation.

The overall agreement between the experiments, simulations, and theory aids in validating our results. Scaling these results to existing simulation work, we observed much lower ion energies[7-11]. This disagreement is directly due to the highly 3D nature of the process. Namely, reducing the process to one transverse dimension constrains the degrees of freedom afforded to the pinch. This was observed when 2D simulations with identical initial conditions to the ones above were preformed[18]. While predicting a lower peak density, these simulations produced a peak proton energy of 13MeV, much higher than what was observed in experiment and 3D simulations. The momentum that would normally go into the second transverse dimension is instead distributed between the other two dimensions resulting in a higher overall kinetic energy. It's interesting to note that the 2D simulations were able to produce the same energy for the on-axis beam. This is due to the longitudinal accelerating force produced by the forward directed electrons.

The presented Z-pinch and acceleration mechanism is directly controlled by the laser and plasma parameters. By tailoring the plasma peak density and gradients it could be possible to explore regions of higher magnetic fields and produce higher density and temperature pinches. This is favorable since the beam current goes as $n_e^{1/3} P^{1/3}$. There is also the possibility of leveraging intense laser ionization to investigate higher Z-materials at ionization states not currently accessible.

## METHODS

Methods and any associated references are available in the online version of the paper.

# ACKNOWLEDGEMENTS

This work was supported by the De**p**artment of Energy and the Naval Research Laboratory Base Program. We would like to acknowledge helpful discussions with B. Hafizi, S. Bulanov, J. Penano, and A. Zingale.

# AUTHOR CONTRIBUTIONS

The experimental apparatus was designed and built by M. H. H. and D. K. with contributes by A. T. The experiments were preformed by M. H. H. and D. K. The samples were processed and analyzed by M.H.H. with the aid of Y.H.C. The numerical simulations were performed by M.H.H. and D.F.G. with analysis code written by M.H.H. M.H.H., D.F.G., A.T., and J.P.P. developed the theoretical model. All authors discussed the results and physical model. M.H.H. wrote the manuscript.

# ADDITIONAL INFORMATION

Supplementary information is available in the online version of the paper. Reprints and permissions information is available online at www.nature.com/reprints. Correspondence and requests for materials should be addressed to M.H.H.

# COMPETING FINANCIAL INTERESTS

The authors declare no competing financial interests.

# METHODS

**CR-39 Analysis**

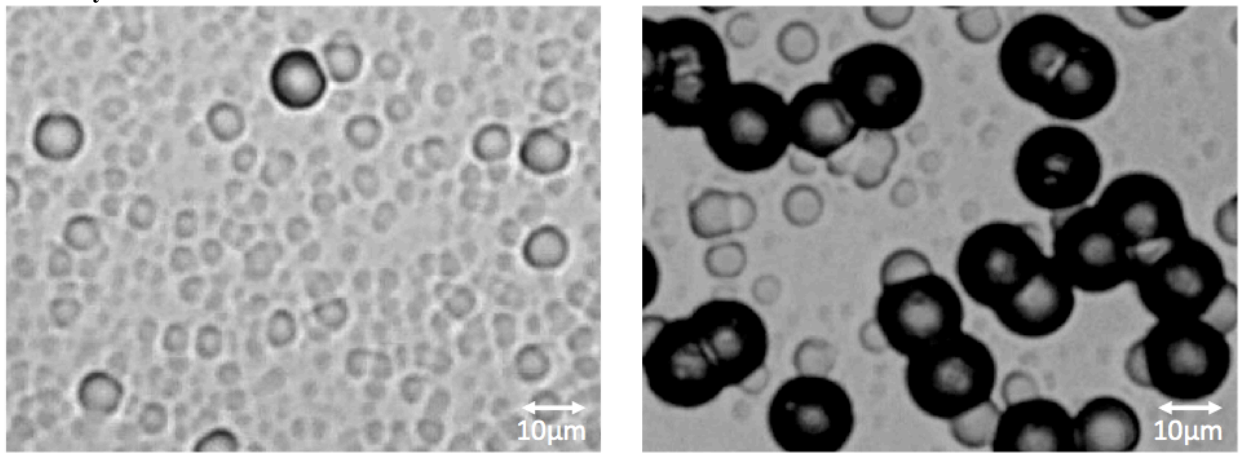

**FIGURE 6.** Microscopic images of etched tracks in CR-39 plates for the 0.6 $n_{crit}$ and 0.3$n_{crit}$ cases.

After exposure, the CR-39 plates were etched following the technique in [1]. The plates were etched in a 6N NaOH solution at a temperature of 80 C for 6 hours in order for the pits to be visible under a microscope. The plates were scanned using an optical microscope with a 10x objective and CCD for capturing images. The CCD resolution is 1280x960 and the imaging field of view is 425x320um. A composite image was formed from these captured images and the pits were counted using the cell counting capabilities of the imaging software ImageJ. Large areas within the unshielded region were saturated with pits and counting could not be preformed, so lower bound estimates were made.

Example microscope images of etched pits for the 0.6 $n_{crit}$ and 0.3$n_{crit}$ are given in Fig 6. Both plates were etched in the same batch and both images are of unsaturated areas within the unshielded region. Based on the measurements of proton energy vs pit diameter given in Ref. 1 for the same etching conditions, we are able to make estimates of the proton energies in the unshielded region when no protons of energy greater than ~800keV are present. For example, in the 0.6 $n_{crit}$ case, there are no pits greater than 10um as well as no observed tracks beyond the 9um aluminum filter (<700keV). This indicates that no protons were accelerated beyond 200 keV. Unfortunately, the results in Ref. 1 do not give a unique solution when proton species with energy both above and below 800keV are present. The proton energies for the unshielded region in the $0.3n_{crit}$ case cannot therefore be extracted.

**TurboWAVE 3D PIC Code**
The underlying acceleration process was simulated using the TurboWAVE fully 3D particle-in-cell simulation code at NRL. The simulation was initialized with the hydrogen gas density profile extracted from the experimental measurements with a peak density of 0.3 $n_{crit}$ and an 800nm 50 fs laser pulse with a focused spot size of 2.6 μm and a normalized vector potential of $a_0 \sim 6$. The pulse is linearly polarized in the x-direction and propagates in the positive z-direction. The numerical parameters were $\Delta x=\Delta y=2.4\omega_p/c$, $\Delta z=0.5\omega_p/c$, and $\Delta t=0.33\omega_p^{-1}$. The simulation included multipass smoothing, ionization, and test particles for tracking proposes. The simulation results were rendered and animated using Paraview.

**Laser Wakefield Scaling Applied to Bennett Pinch**
The temperature of the ions in a Z-pinch can be calculated using the Bennett condition, namely
$$I^2 \mu_0/4\pi = 2 \tilde{n} k_B (T_e+ZT_i)$$
Here I is the electron beam current, $\mu_0$ is the permeability of free space, $\tilde{n}$ is the number of electrons per unit length, $k_B$ is the Boltzman constant, $T_e$ is the electron beam temperature, Z is the ionization state of the ion species and $T_i$ is the ion temperature. The I, $\tilde{n}$, and peak electron energy can be estimated using the scaling laws published in [2,3]. To calculate the beam temperature, we combine the beam energy with the divergence, which was measured separately in experiments (Figure 7). Regardless this value is typically small for laser wakefield accelerators.

We begin by calculating the beam current. As observed in the simulations, a single channel is formed through the entire plasma. This requires the distance that the laser energy depletes, $L_D \approx n_{crit}/n_e \tau_{FWHM} c$, is longer than the dense plasma region, $L_p$. Here $n_{crit}$ is the critical density for 800nm light, $n_e$ is the peak plasma density and $\tau_{FWHM}$ is the laser pulse length. For our experiments $L_D \sim 50$μm and the laser pulse is able to propagate through the target. Additionally, the formation of multiple filaments must be suppressed. This is achieved by allowing the pulse to nonlinearly self-focus within the density upramp; maintaining a matched spot size. If we assume minimal losses to the pulse, by the time it reaches the peak density the normalized vector potential is
$$a_0 = 2 (P/P_{crit})^{1/3} = 2 (P[GW]/17 \, n_e/n_{crit})^{1/3} = 11.6.$$
Here $n_e$ is the peak plasma density and $n_{crit}$ is the critical density for 800nm light. The corresponding blowout radius is then
$$r_B=2 \sqrt{(a_0)}/k_p = 1.5 um,$$
where $k_p$ is the k-vector associated with the plasma wave. If we assume that the bubble becomes fully beamloaded and the electrons are distributed over the length of the bubble, the linear electron density becomes
$$N_{max} = 2.58 \, 10^9 \, \lambda[\mu m]/.8 \, sqrt[P[TW]/100]/2r_B =5.5 \, 10^{14} \, cm^{-1}.$$
This a fair assumption since the dephasing length is much smaller than the plasma, with 1D scaling estimating it to be ~10μm [2].
The maximum attainable energy is
$$\Delta E[MeV] = 38(P/P_C)^{-2/3}P[TW] = 10 MeV.$$
The beam current is then
$$I = e N_{max} \beta c = 26 kA,$$
where $\beta = v/c$.
To solve for the electron thermal energy we take $k_B T_e = \frac{1}{2} \gamma m_e (\beta \sin[\alpha])^2 c^2$, where $\gamma$ is the relativistic Lorentz factor and $\alpha$ is the beam divergence angle. We find $\alpha$ by either using the low energy beam size from the experiment or from image plates exposed during experiments. An example image plate is shown in fig 2. Both the image plate and low energy ions produce beam radius of ~1.5cm at 13cm. This yields an α value of 6.5°. Plugging these values into the Bennett condition and solving for $k_B T_i$, we obtain a ion beam energy of 300keV.

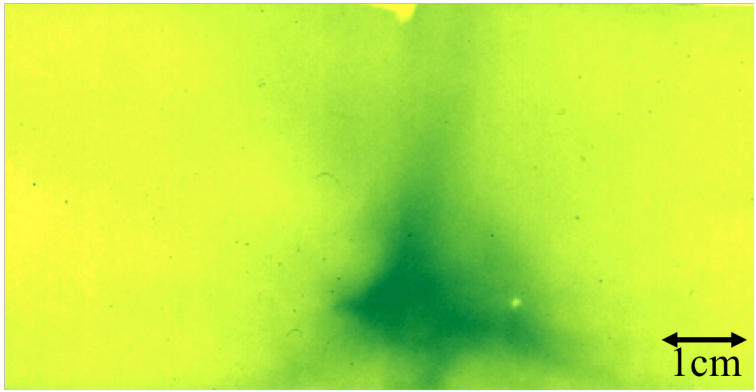

**FIGURE 7.** Electron beam image taken at the detector plane using aluminum shielded GafChromic film. The electron beam radius is ~1.5cm.